\documentstyle[twocolumn,aps,floats]{revtex}

\input psfig.sty
 
\begin{document}
\draft
 
\twocolumn[\hsize\textwidth\columnwidth\hsize\csname %
@twocolumnfalse\endcsname

\bibliographystyle{revtex}


\title{Universal Dynamics of Phase-Field Models for Dendritic Growth} 

\author{
Yung-Tae Kim$^{1}$, Nikolas Provatas$^{1,2}$, Nigel Goldenfeld$^{1}$, 
and Jonathan Dantzig$^{2}$
}

\address{
$^1$University of Illinois at Urbana-Champaign, Department 
of Physics \\1110 West Green Street, Urbana, IL, 61801
}

\address{
$^2$University of Illinois at Urbana-Champaign, Department 
of Mechanical and Industrial Engineering 
\\1206 West Green Street, Urbana, IL, 61801
}

\date{\today}

\maketitle

\begin{abstract} 
We compare time-dependent solutions of different phase-field models for
dendritic solidification in two dimensions, including a
thermodynamically consistent model and several ad hoc models.  The
results are identical when the phase-field equations are operating in
their appropriate sharp interface limit.  The long time steady state
results are all in agreement with solvability theory.  No computational
advantage accrues from using a thermodynamically consistent phase-field
model.
\end{abstract}
\vspace{0.3in}

]

Dendrites are the most commonly observed solidification microstructures in
metals. The free growth of a single dendrite is a prototype for problems
of pattern selection in materials science
\cite{Hua81,Gli84,Lan80,Lan87a,Kes88} and has been extensively studied
experimentally and theoretically.  It is still not possible to compare
theory with experiment because of the difficulties in computing three
dimensional microstructures in the temperature and material's parameter
range of the experiments. 

Recently, a significant step forward was taken by Karma and
Rappel\cite{Kar95,Kar97}, who not only showed how to compute accurately
two dimensional dendritic growth but also were able to compare their
results with theoretical predictions.  Their calculations used the
so-called {\it phase-field\/} formulation of solidification, in which a
mathematically sharp solid-liquid interface is smeared out or
regularized and treated as a boundary layer, with its own equation of
motion.  The resulting formulation, described in more detail below, no
longer requires front tracking and the imposition of boundary
conditions, but must be related to the sharp interface model by an
asymptotic analysis.  In fact, there are many ways to prescribe a
smoothing and dynamics of a sharp interface consistent with the original
sharp interface model, so that there is no unique phase-field model, but
rather a family of related models.  An important part of Karma and Rappel's
work was an improved asymptotic analysis which allows coarser spatial
grids to be used in the numerical computations than was previously possible.

Although the phase-field method has gained acceptance as a useful way to
study solidification problems, a debate still exists over the
interpretation and validity of the phase-field models themselves. Each
model includes a double-well potential field which enforces the above
properties of the phase-field. Some models can be shown rigorously to
satisfy entropy inequality\cite{Pen90,Wan93}. These are sometimes called
``thermodynamically consistent" models.  On the other hand, it has been
argued that the precise form of the phase-field equations should be
irrelevant so long as the computations are performed at the asymptotic
limit where the phase-field model converges to the sharp interface limit
\cite{Lan86a}.

The purpose of this Letter is to compare the dynamics of the different
phase-field models proposed.  To this end, we have performed accurate
and extensive computations using a specially developed adaptive mesh
refinement algorithm\cite{Pro98a,Pro98b}.  We find that when properly
used, all phase-field models give precisely equivalent results;  
not only does each phase-field model converge to the
steady state predicted by theory, but also the transient
dynamics approach the steady state uniquely.  Indeed, once one has
established that there is genuine universal dynamic behavior, the
only remaining consideration is the computational efficiency. Our
results clearly indicate that the CPU times required for the
different models are identical. In particular, we find no advantage
for the thermodynamically consistent model.

A secondary purpose of this Letter is to detect the limit of the
validity of phase-field models in describing the sharp interface
problem.  In the context of the asymptotic analysis of Karma and
Rappel \cite{Kar95,Kar97}, the ratio of the interface width to the
diffusion length (referred to as the interface P\'{e}clet number,
IPe) must be small in order for the different phase field-models to
collapse to identical free-boundary problems. We show how finite-IPe
discrepancies encountered between different models can be eliminated
by adjusting the phase-field parameters. We emphasize that IPe is a
free parameter and can be varied for numerical convenience by
changing the interface width.

The solidification of a pure
substance is described by a free-boundary problem for the temperature
in the solid and liquid phases, and the position of the interface
between them:
\begin{eqnarray}
\partial_t{u} &=& D\nabla^2{u} \\
V_n &=& D(\partial_n{u}|^+ - \partial_n{u}|^-) \\
u_i &=& -d(\vec{n}) \kappa - \beta(\vec{n}) V_n
\end{eqnarray} 
The temperature $T$ has been rescaled as a dimensionless thermal
field $u=(T-T_m)/(L/C_p)$, where $T_m$, $L$, and $C_p$ represent the
melting temperature, the latent heat of fusion, and the specific heat
at constant pressure, respectively. The thermal diffusivity $D$ in
Eq.~(1) is assumed to be equal in both phases. Eq.~(2) describes
energy conservation at the solid-liquid interface, where $V_n$ is the
local outward normal interface velocity and $\partial_n$ refers to
the outward normal derivative at the interface for the solid (+) and
liquid (-) phases. Finally, Eq.~(3) is known as the Gibbs-Thomson condition,
describing the deviation of the interface temperature $u_i$ from
equilibrium, due to the local curvature $\kappa$, and interface kinetics.
$d(\vec{n})=\gamma(\vec{n}){T_mC_p}/{L^2}$ is the anisotropic
capillary length, proportional to the surface tension
$\gamma(\vec{n})$, and $\beta(\vec{n})$ is the anisotropic kinetic
coefficient.

Eqs.~(1)-(3) have been studied extensively to determine the steady
state features of dendritic growth \cite{Lan80,Lan87a,Kes88}.  These
equations admit a family of discrete solutions. Only the fastest
growing of this set of solutions is stable, and this is the
dynamically selected ``operating state'' for the dendrite,
corresponding to a unique tip shape and tip velocity. This
theoretical treatment is usually called solvability theory. Recent
calculations of dendritic growth using phase-field models have been
found to be in good agreement with the predictions of solvability
theory \cite{Kar95,Kar97,Pro98a}.

The phase-field model finesses the computational difficulties
associated with front-tracking on a discrete lattice by introducing
an auxiliary continuous order parameter, or phase-field, $\phi ({\bf
r},t)$ that couples to the evolution of the thermal field. The
dynamics of $\phi ({\bf r},t)$ are designed to follow the evolving
solidification front
\cite{Cag86a,Col85,War95,Kar94,Whe96,Kob93,Wan96}.  The phase-field
interpolates between the solid and liquid phases, attaining a
different constant value in each phase (typically $\pm 1$), with a
rapid transition region in the vicinity of the solidification front.
The liquid-solid interface is defined by the level set of $\phi ({\bf
r},t)= 0$.

We consider phase-field equations of the form
\begin{eqnarray}
&&\frac{\partial u}{\partial t} = D \nabla^2 u + \frac{1}{2} \frac{\partial 
h(\phi)}{\partial t}
\label{phase-field}\\
\nonumber
&\tau (\vec{n})&  \frac{\partial \phi}{\partial t} = \nabla \cdot
(W^2(\vec{n}) \nabla \phi ) - \frac{F(\phi,\lambda u)}{\partial \phi} \\
\nonumber
& + &
\frac{\partial }{\partial x} \left( |\nabla \phi|^2 W(\vec{n})  
\frac{\partial W(\vec{n})}{\partial \phi_{,x}} \right) \\
& + &
\frac{\partial }{\partial y} \left( |\nabla \phi|^2 W(\vec{n})
\frac{\partial W(\vec{n})}{\partial \phi_{,y}} \right),
\end{eqnarray}
as in Refs.~\cite{Kar95,Kar97}. The order parameter is defined by
$\phi$, with $\phi=+1$ in the solid, and $\phi=-1$ in the liquid. The
interface is defined by $\phi=0$. 

The function $F(\phi,\lambda u)=f(\phi)+\lambda u g(\phi)$ is a
phenomenological free energy where $f(\phi)$ has the form of a double-well
potential, $\lambda$ controls the coupling between $u$ and $\phi$, and the
relative height of the free energy minima is determined by $u$ and
$g(\phi)$. The function $h(\phi)$ accounts for the liberation of latent
heat. Anisotropy has been introduced in Eq.(5) by defining $W(\vec{n})=W_o
A(\vec{n})$ and $\tau(\vec{n})=\tau_o A^2(\vec{n})$.  $\tau_o$ is a time
characterizing atomic movement in the solid-liquid interface, $W_o$ is a
length characterizing the width of the interface, and
\begin{equation} 
A(\vec{n}) = (1- 3 \epsilon) \left[ 1 + \frac{4 \epsilon }{ 1 - 3
\epsilon} \frac{(\phi_{,x})^4 + (\phi_{,y})^4 }{| \nabla \phi|^4}\right]
\label{width} 
\end{equation}
with $A(\vec{n}) \in [0,1]$. $\phi_{,x}$ and $\phi_{,y}$ represent partial
derivatives with respect to $x$ and $y$, and the vector
$\vec{n}=(\phi_{,x}\hat{x}+\phi_{,y}\hat{y})/(\phi_{,x}^2+\phi_{,y}^2)^{1/2}$
is the normal to the contours of $\phi$. The constant $\epsilon$
parameterizes the deviation of $W(\vec{n})$ from $W_o$ and is a measure of
the anisotropy strength.

We use the asymptotic relationships of Karma and Rappel
\cite{Kar95,Kar97} to map the phase-field model into the
free-boundary problem, where Eqs.~(4) and (5) reduce to Eqs.~(1)-(3).
In terms of $A(\vec{n})$, $\beta(\vec{n})=\beta_o A(\vec{n})$ and
$d(\vec{n}) = d_o \left[ A(\vec{n}) + \partial_{\theta}^2 A(\vec{n})
\right]$, where $\theta$ is the angle between $\vec{n}$ and the
$x$-axis; noting that $\tan(\theta)=\phi_{,y}/\phi_{,x}$, these
expressions become $\beta(\vec{n})=\beta_o(1 + \epsilon \cos4
\theta)$ and $d(\vec{n}) = d_o(1 - 15 \epsilon \cos4 \theta)$ in the
free-boundary problem. The parameters of the phase-field model are
related to the free-boundary parameters by $\lambda=W_o a_1/d_o$ and
$\tau_o=W_o^3 a_1 a_2/(d_o D) + W_o^2 \beta_o/d_o$. The positive
constants $a_1$ and $a_2$ depend on the exact form of the phase-field
equations. In choosing to simulate particular material
characteristics, we fix the experimentally measurable quantities
$d_o$, $\beta_o$, and $D$, leaving $W_o$ as a free parameter which
determines $\lambda$ and $\tau_o$.

We compute four-fold symmetric dendrites in a quarter-infinite space using
a new finite-element adaptive grid method reported in Refs.
\cite{Pro98a,Pro98b}.  Solidification is initiated by a small quarter disk
of radius $R_o$ centered at the origin. The order parameter is initially
set to its equilibrium value
$\phi_o(\vec{x})=-\tanh((|\vec{x}|-R_o)/\sqrt{2} )$ along the interface. 
The initial temperature is $u=0$ in the solid and decays exponentially
from $u=0$ at the interface to $u=-\Delta$ as $\vec{x} \rightarrow
\infty$, where the far-field {\it undercooling} is
$\Delta=(T_m-T_\infty)/(L/C_p)$ and $T_\infty$ is the temperature far
ahead of the solidification front in the liquid.

\begin{table}[b]
\caption{Summary of phase-field models studied.}
\label{models}
\begin{tabular}{ccccc}
Model & $\frac{\partial g(\phi)}{\partial \phi}$ & $h(\phi)$ & $a_1$ &
$a_2$
\\ \hline
1 & $1-\phi^2$ & $\phi$ & $\frac{1}{\sqrt{2}}$ & $\frac{5}{6}$ \\
2 & $(1-\phi^2)^2$ & $\phi$ & $\frac{5}{4 \sqrt{2}}$ & $\frac{47}{75}$ \\
3 & $(1-\phi^2)^3$ & $\phi$ & $1.0312$ & $0.52082$ \\
4 & $(1-\phi^2)^4$ & $\phi$ & $1.1601$ & $0.45448$ \\
5 & $(1-\phi^2)^2$ & $\frac{15}{8}(\phi - \frac{2}{3}\phi^3 +
\frac{1}{5}\phi^5)$ & $\frac{5}{4\sqrt{2}}$ & $0.39809$
\end{tabular}
\end{table}

\begin{table}[t]
\caption{Summary of simulation parameters.}
\label{sim_params}
\begin{tabular}{ccccccccc}
$\Delta$ & $L$ & $R_o$ & $\Delta x$ & $\Delta t$ & $D$ &
$d_o$ & $\tilde{V}$ & IPe \\ \hline
0.45 & 1000 & 17 & 0.39 & 0.010 & 3 & 0.5 & 0.00545 & 0.011 \\
0.55 & 800 & 15 & 0.39 & 0.016 & 2 & 0.5 & 0.0170 & 0.034 \\
0.65 & 800 & 15 & 0.39 & 0.016 & 1 & 0.5 & 0.0469 & 0.094 \\
0.65 & 800 & 15 & 0.39 & 0.004 & 2 & 1.5 & 0.0469 & 0.031
\end{tabular}
\end{table}                                                                   

The different phase-field models we study are summarized in Table
\ref{models}. To satisfy the asymptotics, $f(\phi)$ is chosen to be
an even function, and $g(\phi)$ and $h(\phi)$ are odd. For all of the
models, $f(\phi)=\phi^4/4 -\phi^2/2$. For computational purposes, the
$g(\phi)$ are chosen such that the two minima of $F(\phi, \lambda u)$
are fixed at $\phi=\pm 1$. Model 1 is a form used by Almgren
\cite{Alm98}, Model 2 is a form used by Karma and Rappel
\cite{Kar95,Kar97}, and Model 5 is the thermodynamically consistent
form used by Wang et al \cite{Wan93}.  Models 3 and 4 are forms
created by us that meet the above requirements.  It should be noted
that Model 1 requires that $\lambda < 1/\Delta$ otherwise the
$\phi=-1$ state becomes linearly unstable.

In our simulations, the computational domain is an $L \times L$ square
box. Computations were performed at $\Delta=0.65, 0.55$, and $0.45$.  A
summary of the parameters used for each simulation is given in Table
\ref{sim_params}, where $\tilde{V}=V d_o/D$ is the dimensionless tip
velocity predicted by solvability theory, $\Delta x$ is the {\it minimum
grid spacing} of our mesh \cite{Pro98a,Pro98b}, and $\Delta t$ is the
simulation time step.  The phase-field parameters were chosen for each
model so that they all simulated the same free-boundary problem. For all
simulations $\epsilon=0.05$, $\beta_o=0$, and $W_o=1$. Figure 1 shows the
dimensionless tip velocity of the dendrite versus time for the simulations
performed at $\Delta=0.55$ and $0.45$. These results show that all of the
phase-field models studied produce identical results for the entire
temporal evolution of the dendrite and also converge to steady state
solutions that are within a few percent of those predicted by solvability
theory. In addition, the CPU times required for each of the models were
identical. 

\begin{figure}[h]
\leavevmode\centering\psfig{file=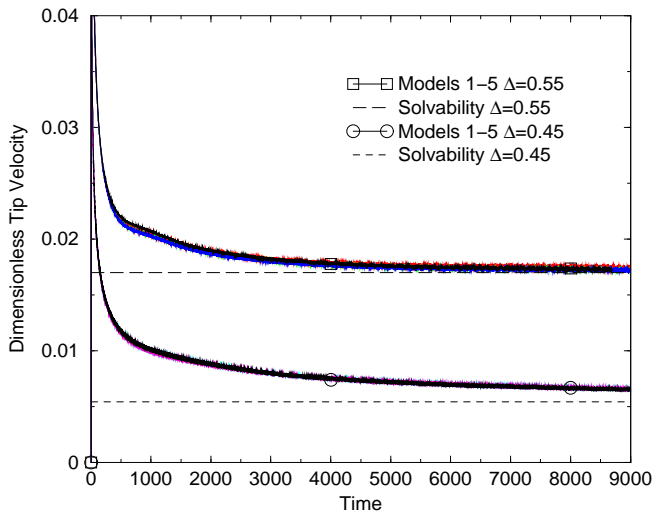,width=\columnwidth}
\caption{Time evolution of the dimensionless tip velocity for five
different phase-field models at $\Delta=0.55$ and $0.45$. Each curve
consists of five solutions superimposed on one another.}
\end{figure}                                                              

At $\Delta=0.65$ (with $d_o=0.5$), however, there are significant
quantitative differences between the various phase-field models, as shown
in Fig.~2. This discrepancy is attributed to finite-IPe corrections at
higher order in the asymptotic expansion. The deviations are a signal that
the solutions have not converged as a function of the expansion parameter
and that the phase-field equations are not operating within the sharp
interface limit. It should be possible to make these higher order terms
negligible if one makes IPe smaller. Figure 2 also shows that each model
has different convergence properties. However, in other simulations we
have found that no single model consistently converges more rapidly than
the others; in general, the convergence appears to depend on the initial
conditions. 

\begin{figure}[h]
\leavevmode\centering\psfig{file=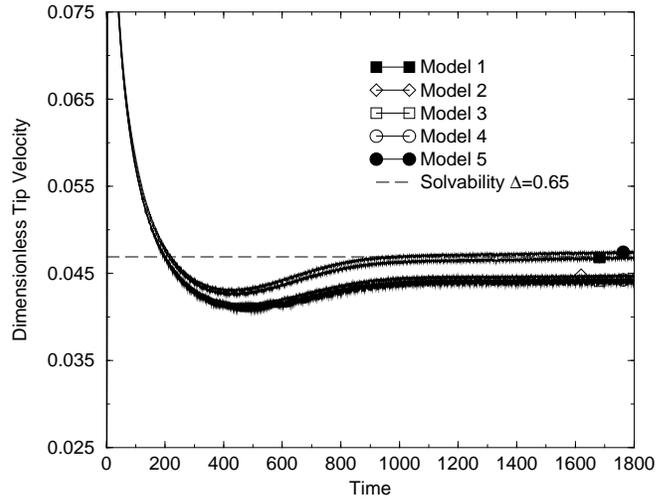,width=\columnwidth}
\caption{Time evolution of the dimensionless tip velocity for five
phase-field models at $\Delta=0.65$ with $d_o=0.5$ and IPe$=0.094$. When
IPe
is too large the different models do not exhibit universal behavior.}
\end{figure}

In their asymptotic analysis, Karma and Rappel expand the solutions to the
phase-field equations treating the interface P\'{e}clet number as a small
parameter \cite{Kar95,Kar97}. This expansion parameter can be written as
${\rm IPe}=WV/D=W_o\tilde{V}/d_o$, where the dimensionless velocity
$\tilde{V}$ depends {\it only} on $\Delta$ and $\epsilon$. We decreased
IPe for the $\Delta=0.65$ case by simulating with a larger value of $d_o$.
Figure 3 shows the tip velocity versus time for the different phase-field
models at $\Delta=0.65$ with the capillary length $d_o=1.5$. The universal
behavior of the different models is recovered after this adjustment is
made. 

\begin{figure}[htb]
\leavevmode\centering\psfig{file=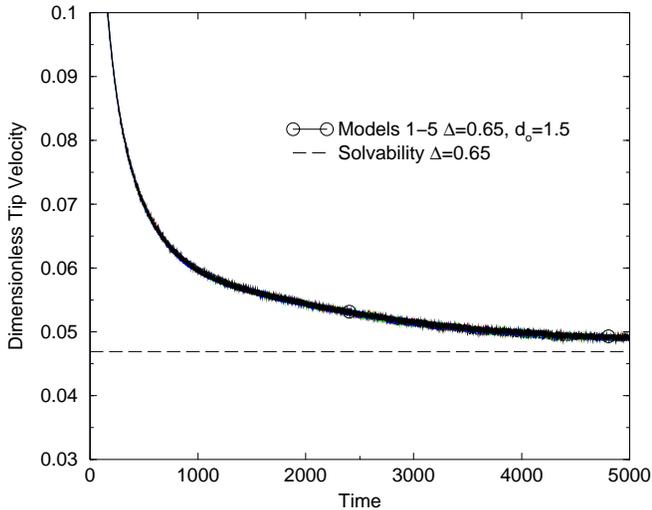,width=\columnwidth}
\caption{Time evolution of the dimensionless tip velocity for five
phase-field models at $\Delta=0.65$ with $d_o=1.5$ and IPe$=0.031$.
Decreasing IPe recovers universal behavior.}
\end{figure}

We have demonstrated that one can obtain identical results from different
phase-field models by choosing the expansion parameter IPe to be
sufficiently small. Unfortunately, in practice, the interface width is the
only parameter that can be used to control the size of IPe, since
$\tilde{V}$ is fixed for a given $\Delta$ and $\epsilon$, and $d_o$ is set
by the particular material to be simulated. Thus, there is only the one
free parameter, $W_o$, that can be adjusted make IPe smaller.  This
restriction can hinder computational efficiency, as the number of grid
points necessary to resolve the interface (and thus the simulation time)
scales as $1/W_o^2$ on an adaptive grid, and as $1/W_o^3$ on a fixed grid.
In addition, with zero interface kinetics, $\tau_o \sim W_o^3$ which
places a restriction on the computational time step if an explicit scheme
is used.  For the simulation at $\Delta=0.65$ with $d_o=0.5$, an IPe
$=0.031$ can be obtained by reducing $W_o$ by a factor of 3, but this
would require an impractical amount of computing time. We note that the
asymptotics of Karma and Rappel become most accurate at lower
undercoolings \cite{Pro98b}, which is also an experimentally relevant
regime.  At low $\Delta$, $\tilde{V} \sim \Delta^4$, allowing the use of
larger values for $W_o/d_o$.  Simulating at low $\Delta$ requires larger
system sizes. This, however, adds very little computational complexity for
adaptive mesh based codes. 

We can extend the range of validity for these phase-field models by
carrying out the asymptotic analysis further so that finite-IPe
corrections are pushed to higher orders. This will lessen the restrictions
on the interface width, thus rendering the phase-field approach
computationally more efficient. Detailed results will be presented in a
forthcoming paper \cite{Hou98}. There appears to be a general trend that
as one goes to higher orders in the asymptotic expansion more constraints
are required on the functions $f(\phi),g(\phi)$, and $h(\phi)$ in order to
get rid of correction terms inconsistent with the free-boundary
formulation. These constraints can cause the phase-field to have solutions
that are not monotonic in the interfacial region, thus requiring higher
grid resolution and computation time \cite{Alm98}. We are currently
pursuing the development of a phase-field model from a renormalization
group approach with the goal of creating a more systematic convergence to
the free-boundary problem.

We thank Wouter-Jan Rappel for providing the solvability code used to 
test some of our simulations, and Alain Karma for generously
providing us with his unpublished results. This work has been supported by
the NASA Microgravity Research Program, under Grant NAG8-1249.  We also
acknowledge the support of the National Center for Supercomputing
Applications (NCSA) for the use of its computer resources.

\bibliography{biblio} 

\end{document}